\documentclass{article}
\usepackage{amsmath,amssymb,graphicx}
\newcounter{rem}
\newcounter{ex}
\newcommand{\rem}{\refstepcounter{rem}\par\emph{Remark \therem.} }
\newcommand{\ex}{\refstepcounter{ex}\par\emph{Example \theex.} }

\newcommand{\karti}[4]{\begin{figure}[#1]\begin{center}
\includegraphics[height=#2]{#3}
\end{center}\caption{#4}\end{figure}}
\newcommand{\ssy}[5]{#1, \emph{#2} {\bf #3}, #5 (#4)}
\begin{document}
\title{The time travel paradox}
\date{}
\author{S. Krasnikov\thanks{Present affiliation:
{\it The Central Astronomical
Observatory at Pulkovo}}\\
\fbox{{\it Starlab NV}}}
\maketitle
\begin{abstract}
We define the time travel paradox in physical terms and prove its
existence by constructing an explicit example. We argue further that
in theories --- such as general relativity --- where the spacetime
geometry is subject to nothing but differential equations and initial data
no paradoxes arise.
\end{abstract}
\section{Introduction}  

Over the last decade the (im)possibility of creating a time machine has been
animatedly discussed in the literature (see \cite{Viss,Nah,phil} for
reviews). The obvious reason is that after a specific (however
unrealistic from the technical point of view) recipe for building a time
machine was proposed in the noted paper
\cite{MT}  the problem of causality
violations drastically changed its status. Until then causality in
classical gravity had been perceived as something like an additional
postulate.  There is no observational evidence that we live in an
acausal (G\"odel's, say) universe, so  the
problem  generally had been dismissed as purely academic. The paper,
however, drew attention to the
fact that within general relativity causality is not  given once
and for all. Even if the Universe is causal at the moment it does
not mean
that it cannot be \emph{made} acausal in the future (by some advanced
civilization).  One might  not have been  interested in
achievements of a hypothetical civilizations, but the very fact was
 worrisome that one could not predict the outcome of a simple
thought experiment (the manipulations with a wormhole proposed in
\cite{MTY}), or rather that the predicted result (appearance of a time
machine) was considered by many as inappropriate.
\par
The attempts to quickly put the jinnee back in a bottle failed. We
understand now that causality can be protected neither by a shortage of
`exotic matter', nor by the instability of the Cauchy horizons, as it
was initially suspected \cite{Conj}. Indeed, exotic matter is
necessary \cite{Conj} for the creation of time machines (TMs) with
compactly generated Cauchy horizons (CTMs), but it was shown that
quantum effects can produce such matter in amounts sufficient to
sustain a traversable wormhole \cite{wh} and hence a wormhole based
time machine.  Besides, CTMs are only a particular type of the time
machines.  No reasons are known (see \cite{HS,phil} and
section~\ref{res-ns}) to consider them as something `better' (more
physical, or more feasible) than the TMs with non-compactly
generated Cauchy horizons (NTMs) while the latter often do not
require exotic matter at all. Similarly, nothing suggests that NTMs
must suffer the instability. And even the CTMs while exhibiting,
indeed, the instability in a few cases, in a few other do not (see
\cite{inst} for a discussion).
There remains, of course, the possibility
 that each (type of the) time machine is impossible by its
own reasons, but it does not look too plausible.
\par
Thus it seems appropriate now to turn attention to the most
controversial issue related to the time machines --- the time travel
paradoxes. On the one hand, these paradoxes seem to be something
inherent to time machines (their main attribute, perhaps), so it is
reasonable to assume that if there exists a universal law prohibiting
the time machines, it must have something to do with the paradoxes.  And
on the other hand, be the problem of paradoxes satisfactorily solved
there probably would be no need to look for such a law, the (supposed)
paradoxicalness of the time machines being traditionally the main objection
against them.
\par
The aim of this paper is to consider the time travel paradoxes in
full detail. Two questions will be our primary concern:
\paragraph{\it 1. Are the time machines associated with any paradoxes in
the first place?} The answer, which once seemed obvious (the famous
grandfather paradox was formulated 70 years ago and is known now
in dozens of versions, see \cite{Nah}), in the last decade has been
getting less and less clear. The point is that an argument like `A
time traveler can kill his grandfather before the former's father
was conceived, but then the traveler would not be born, so nobody
would kill his grandfather, etc.'' while  suggesting, of course, that
time travel may be attended by paradoxes does not \emph{prove} it.
In particular, we have no way of modeling this situation with the necessary
accuracy and so plenty of room remains for all sorts of
resolutions which are abundant in science fiction (traveler
can change his mind, or can kill a wrong person, etc.).
To find a non-trivial paradox thus one needed a much simpler
story, which could have been modeled in such detail as to take into account
all relevant factors. One possibility was the Polchinski paradox
\cite{bill} ---
a ball gets into a time machine, travels to the past and hits its
younger self so that the latter misses the time machine and thus
cannot later hit itself, etc.  A few attempts were made
(see \cite{billiard, RaSe} for example) to construct a specific example of
this paradox. For a
wormhole based time machine \cite{billiard} the problem still turned out to
be too hard mathematically, but for the Deutsch-Politzer spacetime a paradox
eventually was claimed to be found \cite{RaSe}. Later, however, it was
shown \cite{Par} that this paradox is only
\emph{apparent}:  it is not the causality
violation that leads to the  paradox, but
some illegal global condition imposed on the physical content of the model
(see the end of subsection~\ref{defin}).
\par
Thus though it is widely believed that the existence of time machines
must lead to some paradoxes in fact not a single such paradox has
ever been found.
Moreover, in the only model where (due to its exceptional simplicity) the
question has been fully analyzed the paradoxes were found to be
lacking \cite{Par}. So, it might appear that time travel paradoxes in fact do
not exist. In this article we argue that this is not the case.
\par
We start by giving in section~\ref{anal} a definition of the time travel
paradox. This is necessary because a great variety of different things
are called paradoxes in the literature (undeservedly, as a rule).
For example, the terms `person', `one borne by a
woman', or `one whose grandfather was not killed in infancy' are
essentially synonyms. So the question `Why a time traveler (or
whoever else) did not manage to kill the traveler's grandfather?' is
not a bit more meaningful than the question `Why is a ball round?' Still,
after all such `pseudoparadoxes' are ruled out a situation remains
for which the name `paradox' is fully justified. The paradox appears
as the inconsistency (due to local physical laws) of what takes
place in a \emph{causal} region with the fact that the time machine
will come into existence. Thus it is not the grandfather paradox, but
rather a story about a mad scientist who builds a time machine
intending that after it is finished (say, on Thursday) he will commit
a bizarre suicide --- on Saturday he will enter the time machine,
return to the Friday, and shoot his younger self dead\footnote{That
is (in terms of killing grandfathers) the interesting question is
`Why is it that a child who plans, first, to build his time machine
(at the age of 20), then to marry (at 30), and then to force his
offspring to kill him when he is 25, will always fail?' }.
\par
In section~\ref{expex} we build a sufficiently simple toy model (it is a
flat universe populated exclusively by pointlike massless particles)
 and use it to prove the existence of the paradoxes
by constructing a specific example.
\paragraph{\it 2. How to resolve the paradox?}
In this paper we stay strictly within the framework of classical
general relativity\footnote{Correspondingly, we deal neither with
the (quantum) `many-world' theory \cite{Deu}, nor with the
(metanomological \cite{Romero}?) `principle of self-consistency'
\cite{bill}.}.  The evolution of a spacetime in this theory is
fundamentally non-unique. As we argue in section \ref{res-ns} this
non-uniqueness does not let the time travel paradoxes into general
relativity --- whatever happens in a causal region, a spacetime
always can evolve so that to avoid any paradoxes (at the sacrifice
of the time machine at a pinch). The  resulting spacetimes
 sometimes (see example~\ref{mw}) curiously remind one
of the many-world picture.

\section{The time travel paradox}
%
\label{anal}
In this section I specify exactly what  I call the
time travel paradox.  The final definition will be given in
subsection~\ref{defin} after the components of the initial (rough) definition
formulated in subsection~\ref{def-n} are analyzed.

\subsection{Reformulation of the problem in physical terms}
\label{def-n}
A great many seemingly paradoxical situations are discussed in the
literature (one even can encounter the `sexual paradoxes' separated
into a special category \cite{Nah}). It is important for our
purposes, however, that they all amount to the (assumed) fact that
\begin{quote}  \emph{due to the presence of a time machine a system} (an
elastic ball, an armed time traveler, etc.) \emph{has a state}
(the ball moves with a given speed in a given direction, the  traveler
meets his younger self, etc.)
\emph{incompatible with
the laws governing the evolution of the system} (the ball
appropriately deviates when struck, the person kills
whoever he sees, etc.)
\end{quote}
This formulation still needs a lot of refinements of course, but at
the moment  the following should be emphasized.  By replacing the
`circling' part of the story (`the ball hits its former self, so
it does not enter the time machine, so it will not be hit, so it enters,
etc.') with the plain statement that the laws of motion of the ball
\emph{are incompatible}
with its initial state we have not lost anything paradoxical. Indeed,
that mind-boggling circle is nothing more than a proof (by
contradiction) of the incompatibility: `Suppose the ball evolves from
that initial state according to these laws, then it would have
to hit its former self, .... q.~e.~d.'
\par
All known (to me) time travel paradoxes fit in the above formulation.
  Whatever interesting there is in science fiction
 beyond the above statement (a traveler returning to a
world different from what he remembers, or the universe disappearing
in order not to allow of a paradox, etc.), it all relates to
`why is it a paradox?', or `how to cope with paradoxes?', but not
to `what kind of situations should be called the time travel paradoxes?'
In particular, I shall not consider the so-called `bootstrap
paradoxes' \cite{Viss} as a separate class.  Such paradoxes can be
represented, for example, by a story about an engineer, who departs to
the future, reads the patent disclosures on some device, returns to `his' time
 and does patent it \cite{Hein}. Where did the idea of
this invention come from? In another version \cite{Har} the
protagonist receives a note (with a very helpful clue) from his
older self,  keeps this note in his wallet and when (his) time
comes hands the note to the addressee --- to his younger self. Who wrote
the note? Generally, this class comprises the situations in which the
initial data are compatible with the equations of motion, but the
solutions look counterintuitive by whatever reason (often because a
`lion' (see below) is necessary to achieve the compatibility).
Confronting a bootstrap paradox we have a choice: either we find a
specific law which is violated (converting thus the bootstrap paradox into an
ordinary one), or we must admit that the situation --- however
strange it might be --- is not a paradox at all.

\subsection{The requirements on the laws of motion}

An important part of the definition in question are the words `due
to'. Indeed, it seems unreasonable to call the incompatibility of
an initial state with  laws of motion `a time travel paradox'
unless we have at least
\emph{some} reason  to believe that this  incompatibility takes place
\emph{due to}  the presence of a time machine and not just \emph{in} its
presence.  The natural criterion for distinguishing the two
situations would be whether or not arbitrary initial data are
compatible with these laws in a
\emph{causal} world.
\par
Thus, in defining  the time travel paradox it makes
sense to restrict ourselves to the laws that are not `paradoxical' in
the causal worlds. Before formulating the relevant condition, however,
we have to solve a preliminary problem. The point is that generally
one cannot compare
the laws of motion in two different spacetimes (a causal and an
acausal in our case).

\ex \label{neloc} Take the Minkowski plane and make the cuts along
the two spacelike segments $\{t=\pm1, \;-1< x <1\}$. Remove also
the `corner' points $(t=\pm1,x=\pm1)$. Now \emph{preserving their
orientation} glue the banks of the cuts --- the upper bank of the
lower cut to the lower bank of the upper cut and vice versa.  This
surgery results in a spacetime  (see figure~\ref{DP}a) called the
2-dimensional Deutsch-Politzer (DP) time machine. The corner
points cannot be glued back into the spacetime and thus the DP
space has singularities.
\karti{ht}{0.4\textwidth}{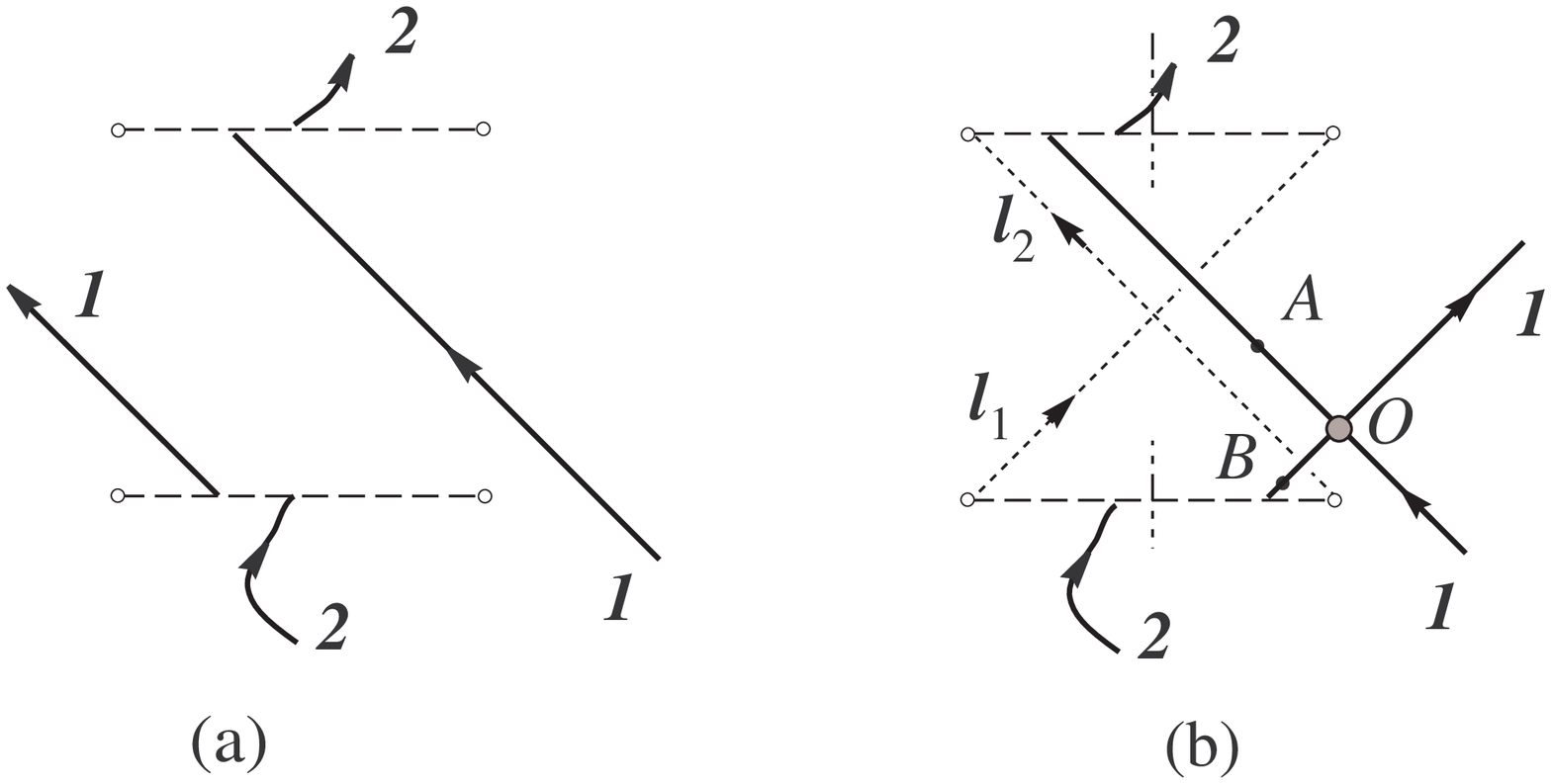}{The
two-dimensional Deutsch-Politzer space (a) and its twisted version
(b). An extension of any continuous curve approaching a (former)
cut must start from the corresponding bank of the other (former)
cut, which looks as discontinuity in the picture. \label{DP}} \rem
In the 4-dimensional case the procedure is the same, but the cuts
are made along the cubes $Q_{1(2)}\equiv\{t=\pm 1,\; -1<x,y,z<
1\}$ instead of the segments. Correspondingly, the singularity in
this case appears as a result of removing the 2-dimensional
boundaries of the cubes (not just  four points). \rem The DP space
is sometimes regarded as something contrived and unphysical (an
attitude which is hard to substantiate within the regular general
relativity, cf.~section~\ref{res-ns}). In this connection note
that the surgery by which we obtained the DP space is just a
convenient way to describe its structure. One can as well define
the Minkowski space as  result of an appropriate surgery applied
to the DP space.
\par
Now consider the two-dimensional Deutsch-Politzer spacetime populated
by particles of two kinds. Let us call them $s$- and $d$-particles
and depict their world lines with single and double lines
respectively. We assign a vector (`momentum') to each particle ---
null to an $s$-particle and timelike to a $d$-particle --- and
require the world line of any particle to be a (segment of) straight
line parallel to its momentum (physically speaking, we describe
a world with two types of particles --- massive and massless --- and
with pointwise interaction).  The laws of motion will be the
following. The world lines can terminate only in vertices.
 Outside a small region $\mathcal R$ around the origin
of the coordinates (the shadowed region in figure~\ref{nelphys}b)
the particles do not interact (an intersection of world lines does
not make a vertex). Inside $\mathcal R$ there are vertices of two
types (see figure~\ref{nelphys}a): (1) if two $s$-particles with
the momenta $\mathbf p_1$ and $\mathbf p_2$ and a $d$-particle
with the momentum $\mathbf p_1+\mathbf p_2$ meet in some point,
all particles colliding in this point annihilate; (2) if among the
colliding particles there are two $s$-particles, but the vertex is
not of type 1 (no $d$-particle with the `correct' momentum) then
the outcome is a single $d$-particle with the momentum $\mathbf
p_1+\mathbf p_2$.
\karti{ht}{0.4\textwidth}{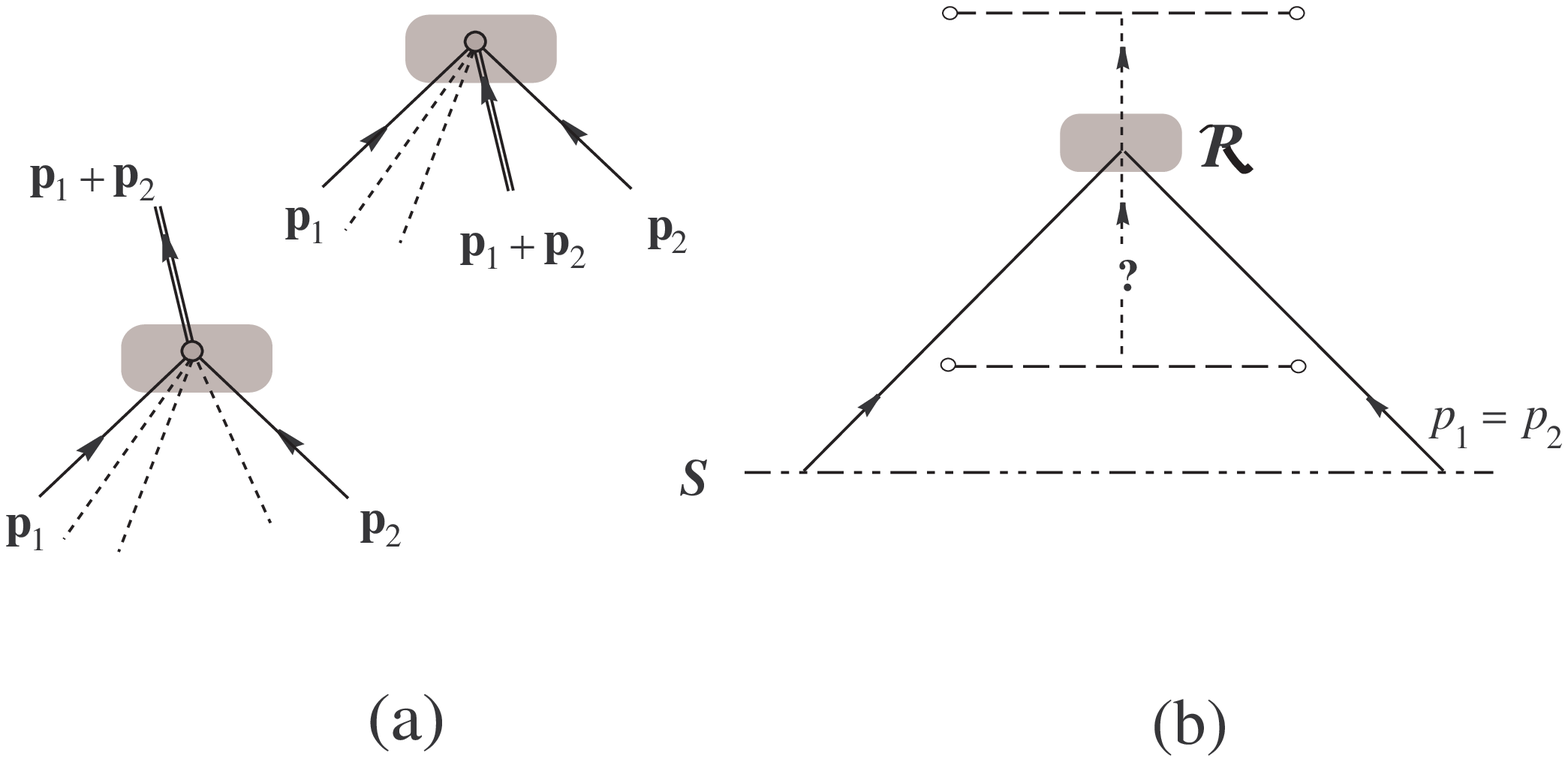}{The local
laws (a) are inconsistent with the initial data  fixed at
$S$.\label{nelphys}}
\par
It is easy to construct a paradox in the described world. For
example, a system of two $s$-particles with the initial data shown in
figure~\ref{nelphys}b obviously cannot obey the laws formulated
above.
\par
We cannot know, however, what
would  happen with these particles in a causal world (the  laws of
motion cannot be transferred to any other spacetime, since, for example, we
cannot distinguish the analog of  $\mathcal R$ there) and so there are no
reasons to attribute this `paradox' to the time machine rather than to
intrinsic
pathologies of the postulated laws of motion.

To avoid such situations it suffices to subject the laws of motion to the
following condition\footnote{Note that in general relativity a much
more restrictive
condition called \emph{local causality} \cite{HawEl} is normally
accepted.}.

\begin{quote}
\textbf{C1 (Locality).} The laws of motion inside any region $U$
do not depend on anything outside of $U$.
\end{quote}
To put it more specifically, (C1) requires that in determining whether
or not a system confined to $U$ obeys a law the answer should
not depend on whether $U$ can be extended to a larger spacetime $U'$
and what are the physical conditions in $U'-U$. Events in $U'-U$ can
influence those in $U$ only via the boundary conditions. The laws
satisfying (C1) we shall call \emph{local}.
\par\smallskip
\rem Condition (C1) rules out, in particular, theories in which
non-locality originates  from incompleteness. One might say, for
example, that the region $\mathcal R$ in example \ref{neloc} is
just a way of description of some field (governed by local laws)
the interaction between particles being dependent on the value of
this field.  Then to make the model complete one would have to
explicitly include the equations of the field in it, which perhaps
would remove the paradox. \rem It is not only the presence of
$\mathcal R$ that contradicts (C1) in the above example. Locality
implies among other things that if some vertex exists in a model
than there must also exist all vertices obtained from it by (in
the flat case) Lorentz transformations and this does not hold in
the example (see figure~\ref{nelphys}a). Note that if we make the
physics of our model local by extending $\mathcal R$ to the whole
spacetime
\karti{ht}{0.4\textwidth}{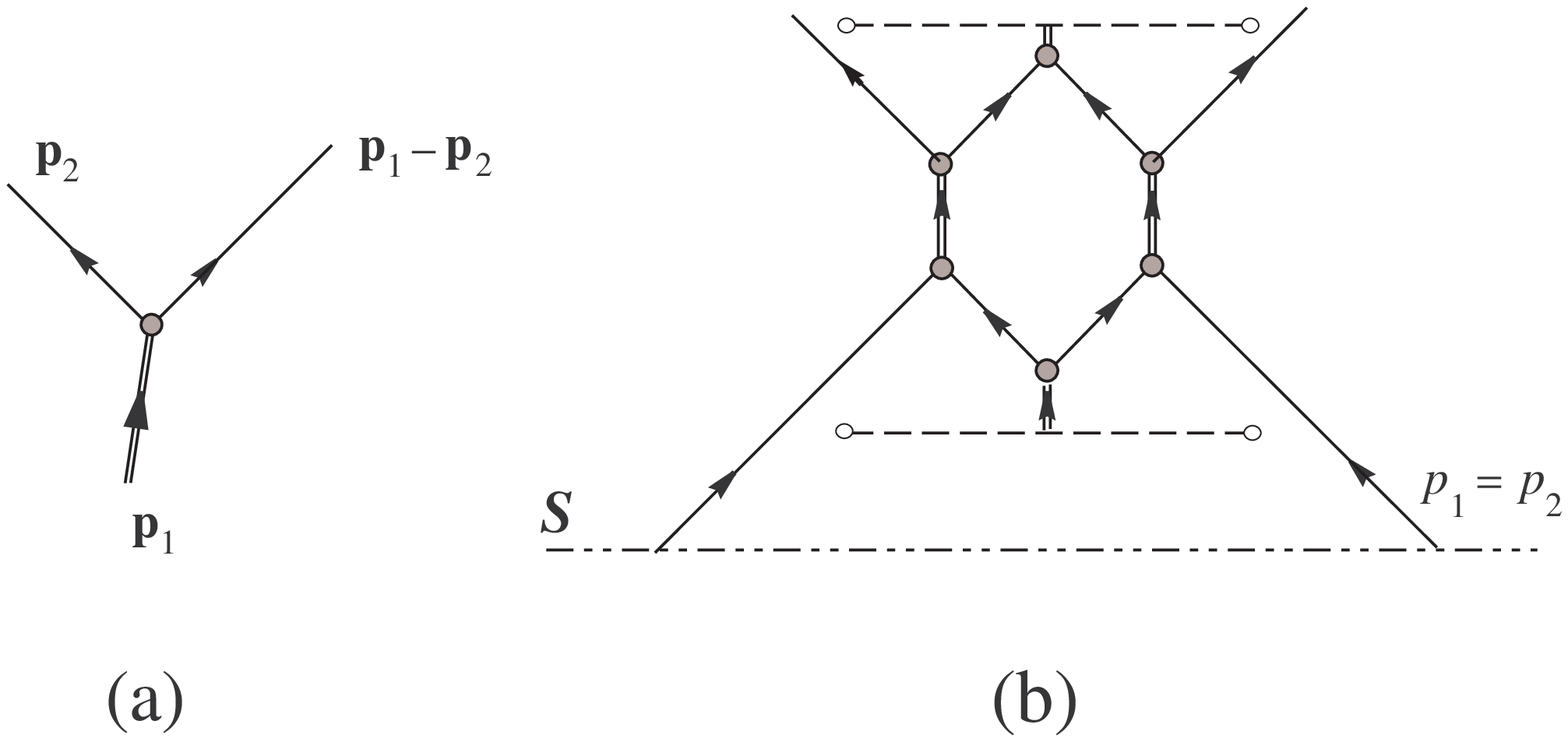}{When the physics of
example~\ref{neloc} is modified so that it becomes local, the
paradox disappears.  \label{neltoloc}} and by adding the missing
vertices (in particular the one shown in figure~\ref{neltoloc}a)
the paradox disappears (see figure~\ref{neltoloc}b).
\par
When the condition (C1) holds the comparison of evolutions in a
causal and an acausal spacetimes becomes meaningful (we always can
cover a spacetime by neighborhoods $\{U_\alpha\}$, where each
$U_\alpha$ is causal) and by the reasons discussed above we impose
yet another condition:
\begin{quote}
\textbf{C2.} In any \emph{causal}
spacetime\footnote{In particular, in any $U_\alpha$, which shows the
non-local nature of the time travel paradoxes.} the laws of motion
must be compatible with any initial data.
\end{quote}
The following example shows that this condition is actually more
restrictive than it might seem. \ex Consider  the massless field
in the Misner space (which is the Minkowski half-plane
$ds^2=-dudw$, $w<0$ with the points identified by the rule
$(u_0,w_0)\longmapsto(\mu u_0,w_0/\mu)$, see figure~\ref{maslfi}).
\karti{h,t,b}{0.46\textwidth}{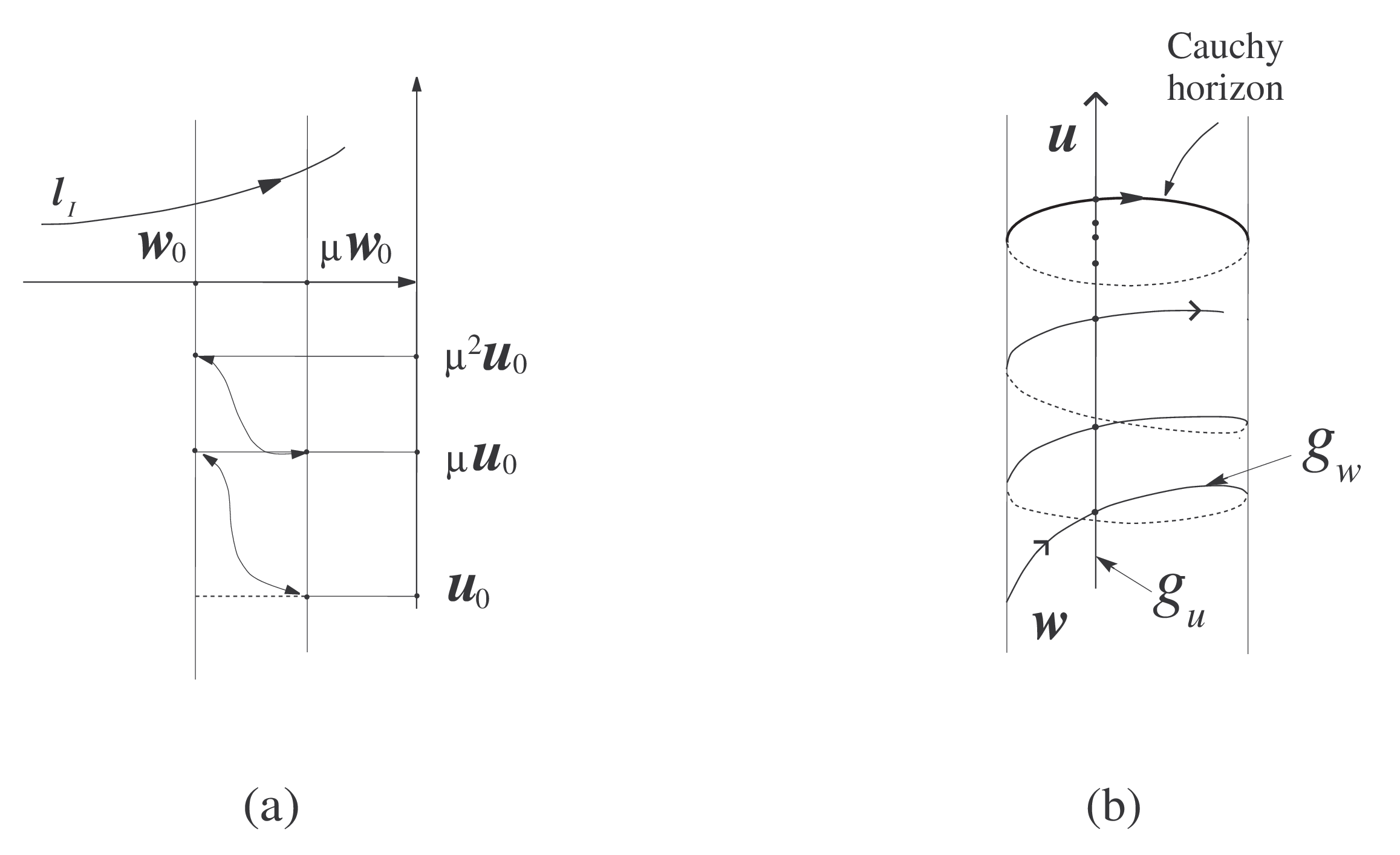}{The Minkowski half-plane (a)
after its points are identified as shown with arrows becomes the
Misner space (b) coordinatized with $u\in \mathrm{I\!R} ,\,
w\in(\mu,1]$. Its lower part ($u<0$) is globally hyperbolic, and
the upper part is acausal.
 \label{maslfi}}
It is easy to show (cf.~\cite{Misdiv}) that the only smooth solution
 in this space of its equation of motion $\square \phi=0$ is $
\phi=const$ [indeed, the null geodesics
$g_u$ and $g_w$ intersect infinitely many times (see
figure~\ref{maslfi}b) before the former reaches the Cauchy horizon, and
each time $\partial_u\phi$ increases by the same factor $\mu ^{-1}$],
which obviously is incompatible with  generic initial data.
However, one can hardly use this situation as a model for a time
travel paradox, since this field may not possess an evolution (for
given initial data) even in causal (though not globally hyperbolic of
course) spacetimes as well. An example of such a spacetime is the
Misner space with a ray $(w=w_0,u\geqslant 0)$ removed from the
acausal part.

\subsection{The lack of information in  causal regions}
In the next subsection we shall return to the discussion of the
definition, but now let us dwell on a fundamental, though
sometimes overlooked property of time machines. The systems in
consideration are classical.  So, intuition (based on what takes
place in globally hyperbolic spacetimes) might suggest that by
fixing their state `at some moment' (that is on a spacelike
hypersurface $S$) we uniquely fix their evolution. But in
non-globally hyperbolic spacetimes \emph{this is not the case}. In
any acausal region  there always are inextendible nonspacelike
curves which do not originate from a causal region. Those are the
closed curves (e.~g.\ $l_L$ in figure~\ref{lions}) and the curves
intruding the spacetime `from nowhere' (e.~g.\  $l_n$ in
figure~\ref{DP}b and $l_I$ in
figures.~\ref{maslfi}a,~\ref{lions}).
\karti{ht}{0.4\textwidth}{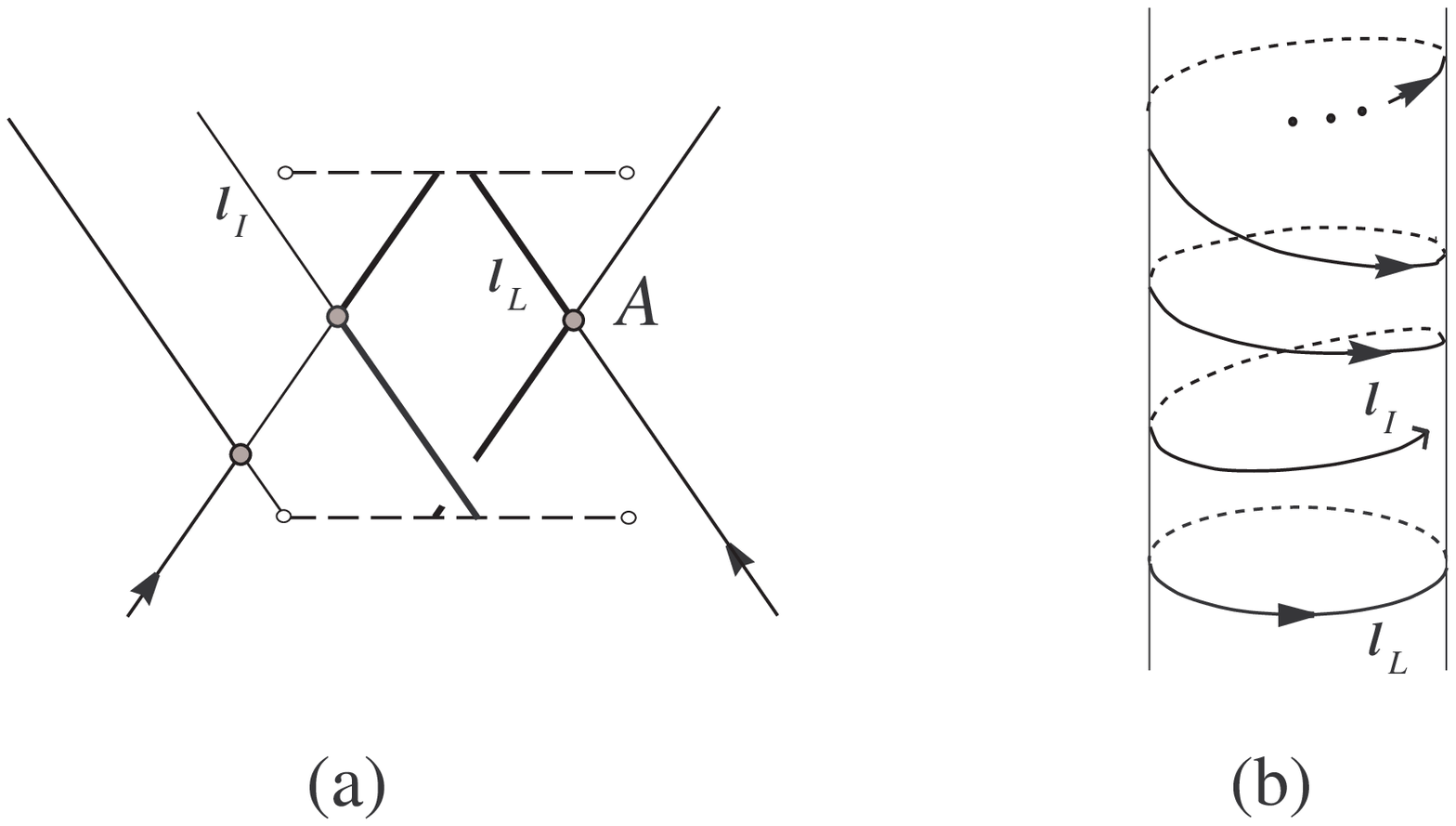}{Different types of lions in the
DP (a) and Misner (b) spaces. $l_L$ loop in the acausal regions
and $l_I$ intrude the spacetimes from a singularity (a), or from
infinity (b). (The latter is the same object as that in
figure~\ref{maslfi}a.) \label{lions}} Along such curves some
`extra unpredictable' (to an observer getting into the time
machine from the causal region) information  can enter the
spacetime (cf.~\cite{Conj}). These curves may be the world lines
of some particles that neither existed prior to the time machine,
nor originated from only the `pre-existing' particles. The unusual
origin of the `unexpected' particles does not make them any less
physical than the conventional ones (which, after all, also could
appear only from either infinity, or a singularity) and so in our
considerations we must take the particles of both kinds equally
serious. The more so, as the line between `looping'
particles\footnote{Aka `self-sufficient loops' \cite{Har}, `jinn'
\cite{jinn}, `self-existing objects' \cite{Romero}.} $l_L$ and
`normal' ones is not always drawn absolutely clear. For example,
$l_L$ in figure~\ref{lions}a can be interpreted as a normal
particle emitted in $A$ and later (by its clock) absorbed there.
\par
The fact that entering a time machine one can meet there an object
whose existence it was impossible to predict is of great
importance not only for resolutions of particular paradoxes (an
unexpected hungry lion behind the door of a time machine could
effectively reconcile the traveler's freedom of will and his
grandfather's safety), but for the very concept of the paradox.
Indeed, most if not all of paradoxes proposed in the literature
are based on some (implicit, as a rule) assumptions about presence
of looping or intruding objects --- from now on I shall use
abbreviation \emph{lions} for them --- and their properties. For
example, the puzzling piece of paper figuring in \cite{Har} (see
subsection~\ref{def-n}) is a typical lion. Another example
(cf.~\cite{RaSe}) is the paradox resulting from the assumption
that \emph{no} lions at all will appear in the acausal region of
the DP space when a pair of elastic balls is prepared in the
initial state shown in figure~\ref{lions}a  by arrows (see
\cite{Par} for more details).
\subsection{The definition of the paradox}
\label{defin} A possible way to obviate the problems with
uniqueness of evolution would be to fix the `initial data' (they
are not  initial data in the ordinary sense as we argue below) for
all relevant particles \emph{including} lions. In the case of the
DP space for example, it could be done (see figure~\ref{grfath}a)
by choosing the surface $t=0$ to be the `initial surface'. This
must be done with caution, however. Such a surface is not
\emph{achronal}, i.~e.\ some points in it can be connected by
timelike curves (that is why  data fixed there cannot be
rightfully called initial). To be consistent with the evolution
laws these data must satisfy some constraints even in theories
free from any paradoxes. \ex \label{tr} Consider the DP space
populated by stable non-interacting particles. The `initial data'
shown in figure~\ref{grfath}a (a single particle moving at $t=0$
to the right) are inconsistent with any possible evolution. To
obtain a `paradox' of precisely the same nature in a globally
hyperbolic spacetime take the surface $t=\varphi/2$ as the initial
surface in the cylinder $t\in \mathrm{I\!R} ,\, \varphi\in(0,1]$
with the metric $ds^2=d \varphi^2-dt^2$, (see
figure~\ref{grfath}b).
 \karti{ht}{0.4\textwidth}{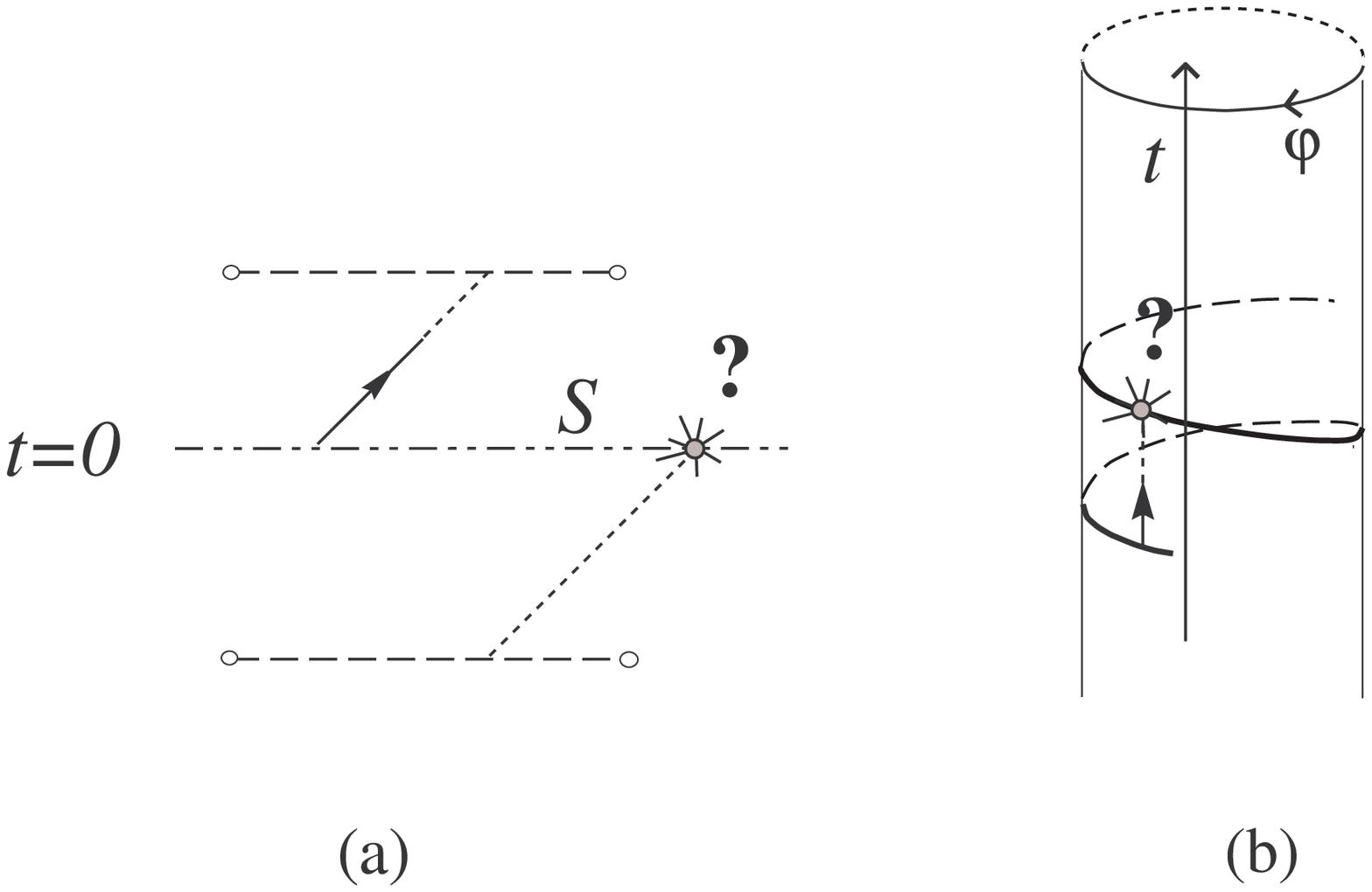}{Trivial
paradoxes in an acausal (a) and in a globally hyperbolic (b)
spacetimes. No evolution of a freely moving particle is consistent
with the initial data depicted by the arrows.\label{grfath}}

To exclude such trivial `paradoxes' we (at the sacrifice of
uniqueness of evolution) shall fix the initial data in the proper way
--- at achronal surfaces (and thus outside of the acausal region).
Correspondingly, we adopt the following final definition of the
paradox:
\begin{quote}
 \textbf{Definition.} The paradox is the inconsistency of some laws
of motion satisfying conditions (C1,C2) with initial data fixed in a
causal region.
\end{quote}

It should be stressed that in doing so we do not miss anything really
paradoxical. Indeed, suppose we find that a time traveler  inside a
time machine is prohibited from doing something (say, from killing a
younger person he meets there)  that he would be
allowed to do  if exactly the same situation (the traveler in
the same mood embedded in exactly the same local environment) took
place in a causal region. Not observing any mechanism that could
enforce the prohibition one might want to call the situation
paradoxical.

Consider, however, an adventurer who wants to build such a time
machine and to try to violate the prohibition --- to kill his younger
self (just in order to find out what can prevent him from doing
this).  The adventurer, when he is in the causal region
yet, is free to make any preparations he wants.
He can choose whatever good weapon, he can adopt any
strategy for his behavior inside the time machine, etc. --- lost labor.
There always will remain an
evolution (i.~e.\ a history not containing the paradoxical suicide)
consistent with all these preparations (or, otherwise there would be a
paradox in the sense of the above definition).
What will be the immediate cause of his failure (whether he will be
eaten by a lion, or just miss the target) depends on the details of
the situation and is immaterial (as long as no new local physics is
involved, which is supposedly the case). What counts is that he is allowed to
make his attempt and so his life does not cost him his freedom
of will\footnote{This reasoning does not work if the adventurer is a
lion, but I do not think that a lion's freedom of will should be
anybody's concern.}. Thus it would be unjustified to call such a
situation a paradox.

\rem The `trivial
paradoxes' (see example~\ref{tr}) that we ruled out  by adopting
our definition constitute the vast
majority of what is traditionally considered as
paradoxes (these are the grandfather paradox and its numerous
modifications). Still our understanding of what is the time travel paradox
is not new: `\dots if there are closed timelike lines to the future
of a given spacelike hypersurface, the set of possible initial data
for classical matter on that hypersurface can be heavily
constrained compared with what it would be if  the same hypersurface
with the same local interactions were embedded in a chronology-respecting
spacetime' \cite{Deu}.

\section{A paradox}
\label{expex}
\subsection{The geometry}
The simplest model of the time machine --- the DP
space --- unluckily for our purpose does not harbor paradoxes (at
least when the  physics is simple enough) \cite{Par}. So for
a paradox we have to choose a spacetime with a
slightly more complex geometry.
\par
Perform the same manipulations as in
constructing the DP space (see example~\ref{neloc}), but this time
glue the upper bank of the lower cut to its counterpart only after it
is \emph{reflected} with respect to the $t$-axis. In other words a
point with $x=x_0$ of this bank is now glued to the point with
$x=-x_0$ of the other bank (not to the point with $x=x_0$ as it was
in the DP case).  The resulting spacetime --- let us call it
\emph{the twisted Deutsch-Politzer (TDP) space} --- is
non-orientable\footnote{The TDP space is in fact the DP space with a
cylinder replaced by a M\"obius strip.} though still time-orientable,
of course. Note that in the TDP space a null geodesic (see curve 1 in
Fig~\ref{DP}b) entering the time machine always has a self-intersection.
\subsection{The physics}
\label{phys}
The world in our model is populated by massless pointlike particles
(that is particles moving
along  null geodesics that terminate only in vertices). Two parameters
are assignd to each particle --- the `color' ($c$) and the `flavor' ($f$) with the
possible values
$$
c=b,g,r\qquad f=\pm1
$$
We write $c^f$ (or $g^{-1}$, $b^1$, etc.) for a particle with the
color $c$ (respectively, $g$, $b$) and the flavor $f$
(respectively, 1, -1).
The particles do not  interact  (an
intersection of their world lines does not make a vertex), with a
single exception --- when two particles
of the same kind (i.~e.\ of the same color and flavor) meet, they
change their flavor
 (see figure~\ref{physpar}).
\karti{ht}{0.3\textwidth}{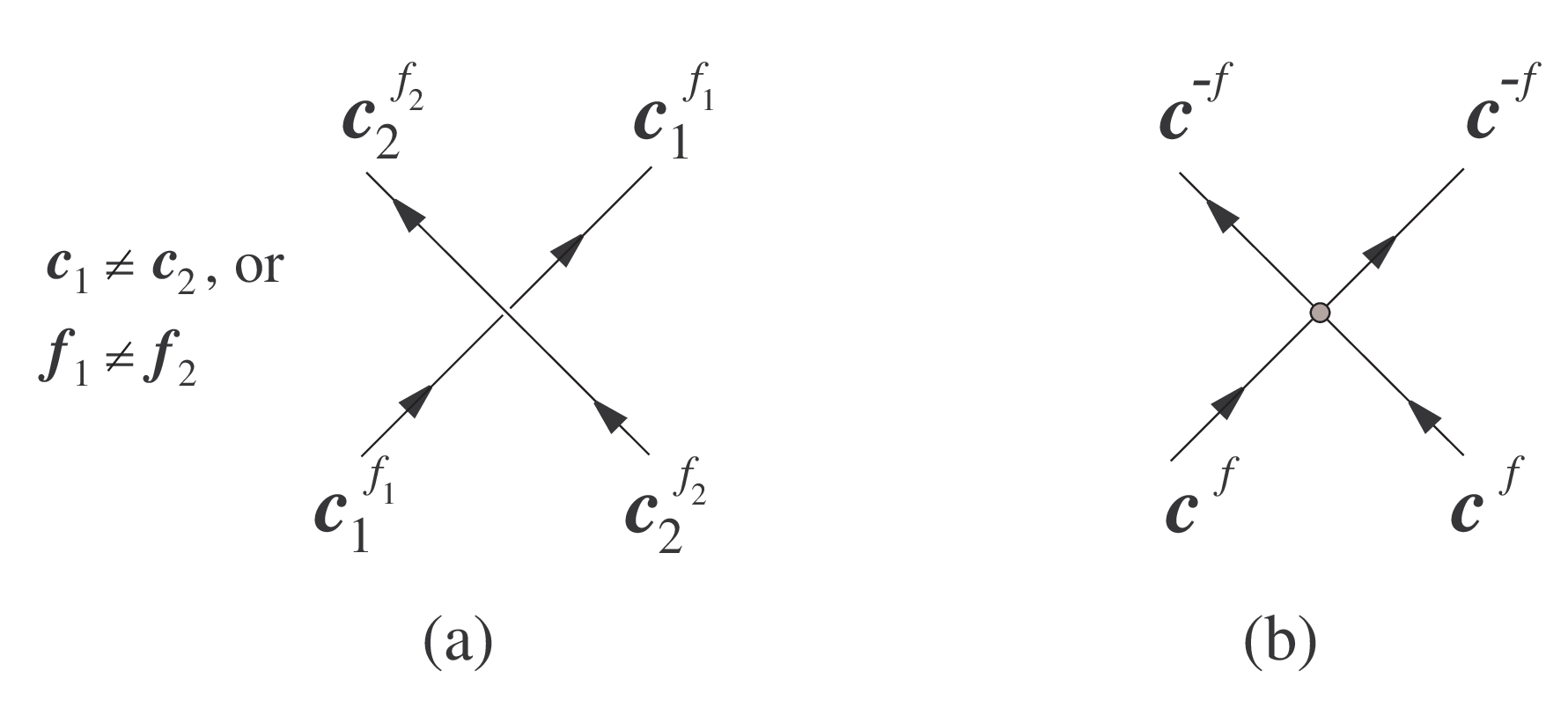}{\label{physpar}The laws of
motion guarantee that in pair collisions the left outgoing
particle differs from the left incoming one.} \rem If required,
one can adopt another (though equivalent in the case at hand)
point of view and speak not  about penetrable  particles, but
about particles that bounce from each other with their parameters
changed according to figure~\ref{physpar}.
\subsection{The paradox}
Let us consider  the system with the initial data posed as shown
in figure~\ref{paradox}: at some moment (the surface $S$)
preceding the appearance of the time machine there are three
particles --- of three different colors --- moving so that they
must get into the time machine.
\karti{htb}{0.35\textwidth}{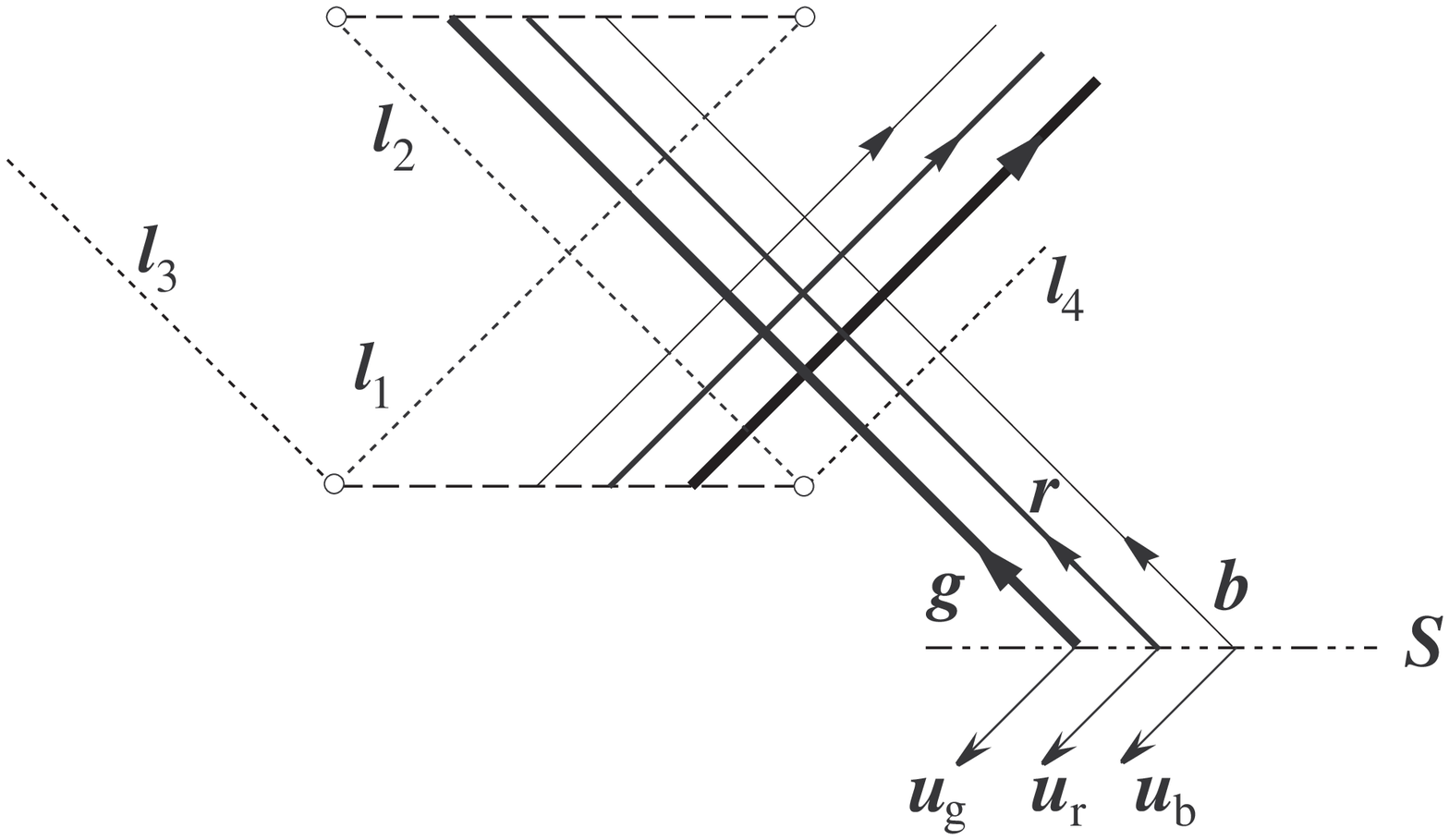}{\label{paradox} All possible
trajectories are shown kinematically compatible with the data
fixed at $S$.}
\par
These initial data do not allow any  evolution,  indeed. To see this
consider the would-be trajectories of the particles after their  collision
with the lion $l_4$. There remain
only two lions ($l_1$ and $l_2$) that can be met on their way. So,
there will be at
least one particle  --- let it be the red one ($c=r$) for
definiteness --- which will not collide any more with a lion of the
same color:
$$
c(l_1)\neq r \neq c(l_2)
$$
Its world line (if it existed)  would be the geodesic 1 in
figure~\ref{DP}b.  It is easy to see that no flavor can be assigned to
the particle on the segment $OA$.  Indeed, on the (open) segment
$(OABO)$ of its world line it does not experience
collisions with other red particles and hence its flavor
cannot change. Hence the flavor -- along with the color --- on the
segment $BO$ must be the same as on $AO$, which is prohibited by the
local physics (see figure~\ref{physpar}b).
\rem Insofar as we discuss \emph{penetrable} particles it is not always possible to assign a particular color to each world line. Two or three particles with the different colors may have the same world line (it can be, for example, any of the geodesics $l_i$ in figure~\ref{paradox}) forming thus a `composite' particle. Such a particle, however, will be sterile. According to the laws specified in subsetion~\ref{phys} interaction occurs only when \emph{two} particles (with the same colors) meet and hence a composite particle passes through any other particle (composite, or not) causing no changes.
\subsection{A four-dimensional version}
The paradox constructed in the previous subsection satisfies all our
requirements. However, it has two (slightly) objectionable features.
First, the spacetime is non-orientable. Second, the transformation $x
\leftrightarrow t $
though not being an isometry still strongly resembles it. So, appealing to
 condition (C1) one might argue
that the existence of the vertex $c^{f} + c^{f} \to
c^{-f}+c^{-f}$ (see figure~\ref{physpar}b) should imply the existence
of the vertex $c^{f}+c^{-f} \to c^{f} + c^{- f}$, which would
destroy the paradox.  We shall show now that in the four-dimensional
case both these `flaws' can be eliminated.
\par
To build the spacetime that we shall use, repeat the construction of
the four-dimensional DP space as is described in the beginning of
section~\ref{def-n}, but before gluing the cubes $Q_1$ and $Q_2$
rotate one of them by $\pi$ in the $(x,y)$ plane. The resulting
spacetime $M$ is orientable and the transformation $x \leftrightarrow
t $ does not even resemble an isometry any longer. At the same time
the two-dimensional section $(y=z=0)$ of $M$ is exactly the TDP
space.
\par
We cannot construct a paradox this time by simply using the same
initial conditions and  physics as in the two-dimensional case.
The singularities now are formed by the (former) boundaries of the
cubes. So, there may be infinitely many lions colliding with the
initial particles, in contrast to the 2D case where there were
only 3 such lions. To handle this problem we shall slightly modify
the local physics. Namely, in addition to color and flavor we
assign to each particle a null vector $\mathbf u$ which parallel
propagates along the world line of the particle and
 does not
change in collisions. We require that particles 1 and 2 should not
interact unless
\begin{equation}
\label{unl}
\mathbf u_1 \parallel \mathbf v_2,\quad \mathbf u_2 \parallel \mathbf v_1,
\end{equation}
where $\mathbf v_i$ is the four-velocity of the $i$th particle, and
when \eqref{unl} holds they should interact exactly as in the
two-dimensional case.
\par
The so chosen laws guarantee that a particle with $\mathbf u$ lying
in the $(t,x)$ plane can interact only with the lions whose world
lines also lie in this plane. Thus restricting consideration to this
plane and choosing $\mathbf u$ for the three initial particles as is
shown in figure~\ref{paradox} we reduce the problem to the two
dimensional one and obtain a paradox.

\section{The paradox as an argument against  time machines}%
\label{res-ns}

In the previous section we proved that the problem (the time travel
paradox) really exists. Now let us discuss how to solve it.\par
Three things --- the set of local physical laws, the initial data, and
 the causal structure of the spacetime --- individually (supposedly)
allowed, turned out to be in conflict.  Correspondingly, (at least)
three ways out are seen.
\par
The system in our toy paradox was governed by exceedingly simple
laws. One might conjecture that the paradox is just a result of this
simplicity while the real (much more rich) physics is free from paradoxes
and any initial data are admissible irrespective of the causal structure.
It is impossible to \emph{refute} (or to prove, for that matter) this
conjecture, but, on
the other hand, it is not obvious why a more complex and detailed theory must
lead to fewer rather than to more paradoxes.
\par
Also one can just \emph{ignore} the paradoxes. After all only
\emph{some}
(but not all imaginable) initial conditions are realized in the
Universe. So, one can argue that the question `Why this or that
particular situation is forbidden?' is senseless, while the `correct'
question is `Whether this situation is realized in the (unique)
existing Universe?' All complications arising then with the notion of
free will is a problem rather philosophical than physical. In my
opinion, however, this approach is somewhat too universal to be
interesting.

Finally, the third possibility is to question the feasibility of the
spacetime geometry involved in the paradox. Instead of asking whether
\emph{time travel} is associated with paradoxes one can ask whether
\emph{general relativity} is associated with them. The difference
between the two questions is that in answering the latter we must
consider the geometry of the universe not as a given background, but
rather as a part of the system, on the same terms as the would-be time
traveler. Correspondingly a paradox now must be defined as
an inconsistency of some initial data fixed (in a causal region) for
a material system \emph{and} a spacetime with their laws of
evolution.  By the `laws of evolution of the spacetime' we understand
(as we restricted ourselves to classical gravity) the postulate
according to which the Universe is described by a maximal (that is inextendible) time-orientable spacetime, the latter being defined as a smooth
four-dimensional Hausdorff manifold with the Lorentz\-ian metric
obeying the Einstein (or other similar) equations.
\par
The two definitions of a paradox are equivalent only as long as we
neglect any influence of matter on the geometry and also believe in
uniqueness of evolution of the spacetime.

That the evolution of a spacetime is not fixed uniquely by initial
data and that due to this fact the existence of a time travel paradox
does not necessarily entail anything paradoxical from the
relativistic point of view can be immediately exemplified by the
paradox constructed in the previous section. Indeed, the initial data
(including the geometrical part) fixed at $S$ are, in fact, quite
\emph{compatible}
with the laws of motion (though not with all possible evolutions
agreeing with these laws). In particular, the spacetime may evolve
just in the Minkowski space. The three particles will fly unimpeded
in this space and no paradoxes will arise.
\par
It may appear that so can be resolved  only paradoxes involving  some
special types of the time machine (e.~g.\ the time machines with
the non-compactly generated Cauchy horizons\footnote{This illusion is
probably the reason why such time machines are sometimes regarded as
 `less physical' than CTMs \cite{Conj}.}).
Indeed, the whole concept
of the time machine as something made by human beings (as opposed to
something appearing `spontaneously' like the Deutsch-Politzer time
machine) is based on the assumption \cite{MTY} that one can
\emph{force} a spacetime to evolve into a time machine (cf.~`the
potency condition' of \cite{phil}), or to put it otherwise that there
exist such initial conditions that a spacetime evolving from them
would inevitably produce a time machine.  However, this assumption is,
strictly speaking, utterly groundless within the limits  of pure
(i.~e.\ without any additions to the above postulate) general
relativity. In this theory \emph{no spacetime at all} evolves
uniquely. Whatever extension $M'\varsupsetneq M$ one takes to be a
possible evolution of the initial spacetime $M$, there always exists
another spacetime $M''$ --- infinitely many such spacetimes, in fact
--- which also presents a possible evolution of $M$. For example, one
can remove a 2-sphere $\Sigma$ from $M'-M$ and take the universal covering
of the resulting space $M'-M -\Sigma$ as $M''$ \cite{chru,HS}. Note, in
particular, that locally $M'$ and $M''$ are isometric and thus
neither of them is preferable from the point of view of the Einstein
equations\footnote{It is often said that the non-uniqueness of
evolution is due to the fact that the Einstein equations do not fix
the \emph{topology} of a spacetime. That is the truth, but not the
whole truth. Pick another sphere $\Sigma'$ and consider the universal
covering $M'''$ of $M'-M -\Sigma'$.  $M'''$ has the same topology as
$M''$, but (in the general case) a different geometry.}.
\par
To overcome such a disastrous lack of predictability one usually
restricts one's consideration to some particular class of spacetimes,
that is in effect one introduces a new (additional to that formulated
above) postulate in the theory.  Not infrequently it is something
like `a spacetime must remain globally hyperbolic as long as
possible' (among the other possibilities is the requirement that a
spacetime should be hole-free). Though incorporation of this
postulate (after appropriate refining) could give rise to an
interesting theory we shall not consider it in the present article
because, anyway, such an `improved relativity' fails at the Cauchy
horizons that bound (as in the case of the wormhole-based time
machine, or the Misner space)
\emph{maximal}
Cauchy developments. A spacetime has infinitely many extensions beyond
such a horizon and \emph{none} of them is globally hyperbolic.

As there are no grounds in the theory to prefer a particular class of
extensions the only remaining way to construct a paradox would be to
find such initial data in the causal region that \emph{all} possible
evolutions would lead to formation of a time machine in the
future. This, however is impossible.  The following theorem can be
proved \cite{th}: \emph{any} spacetime $M$ has a maximal
extension containing no closed causal curves  except perhaps
those lying in the past of $M$.

\ex Let $M$ be the causal part of the Misner space (see
figure~\ref{maslfi}b). There are two
\label{mw} well-known `natural' ways \cite{HawEl}
 to extend it to the whole Misner space
(i.~e.\ to a cylinder with closed timelike curves in it). But there
are also  infinitely many ways to extend it to  a maximal
\emph{causal} spacetime. For example (cf.~\cite{phil}), let
$M'$ be the Misner space with the ray $(w=w_0,u\geqslant 0)$ cut out
from the acausal part .  Take a sequence $\{M'_n\}$, $n=\dots,-
1,0,1\dots$ of copies of $M'$. For each $n$ glue the right bank of
the cut in $M'_n$ to the left bank of the cut in $M'_{n+1}$.  The
resulting spacetime is a  causal maximal  extension of $M$.

If desired (see \cite{Viss2}) the abovementioned theorem  could serve
as basis for postulating causality. What is more pertinent to the
present paper, it proves that \emph{there are no time travel
paradoxes in general relativity}: for any system governed by laws
satisfying (C1,C2) any initial data are allowed in a causal region of
the universe.

\section*{Acknowledgments}
I am grateful to I. Raptis and R. R. Zapatrin for giving me an opportunity
to discuss this paper with them.

\end{document}